\documentclass[12pt]{article}%
\usepackage{amsmath}
\usepackage{amsfonts}
\usepackage{amssymb}
\usepackage{graphicx}%
\providecommand{\U}[1]{\protect\rule{.1in}{.1in}}
\providecommand{\U}[1]{\protect\rule{.1in}{.1in}}

\begin{document}

\title{Generalized forms and gravitation}
\author{D C Robinson\\Mathematics Department\\King's College London\\Strand, London WC2R 2LS\\United Kingdom\\email: david.c.robinson@kcl.ac.uk}
\maketitle

\textbf{Abstract:} \ The algebra and calculus of generalized differential
forms are reviewed and employed to construct a broad class of generalized
connections and to investigate their properties. \ Ths class includes
connections which are flat when Einstein's vacuum field equations are
satisfied. \ Generalized Chern-Simons action principles are formulated and it
is shown that certain of these have Einstein's vacuum field equations as
Euler-Lagrange equations.

\newpage

\section{Introduction}

\ Generalized differential forms have been employed in a number of different
geometrically\ and physically interesting contexts. \ These include twistor
theory, the construction of action integrals and the study of vector fields
and path space forms, \cite{sp1} - \cite{lah2}. \ The aim of this paper is to
review aspects of the algebra and calculus of generalized differential forms,
as developed in \cite{rob1}-\cite{rob8}, and to study a class of generalized
connections that can be employed in the study of gravitational systems and
action integrals.

A review of the basic ideas of the algebra, differential and integral calculus
of generalized forms that are needed here is given in the second section.
\ The third section contains a discussion of generalized connections and
generalized Chern-Simons integrals are constructed. \ \ Certain of the
formulae in this section are similar to those which arise in the formalism of
higher gauge theories reviewed in \cite{baez} although the approach taken
here, based on \cite{rob5} and \cite{rob6}, is different. \ The fourth section
deals with connections for gravitational fields without matter. In particular
generalized connections with values in the Lie algebras of special orthogonal
groups are discussed and flat connections are related to gravitational field
equations. \ The generalized equations of parallel transport for these flat
connections constitute a system of linear equations with these gravitational
equations as integrability conditions. \ Section five deals with the
construction of generalized Chern-Simons three-forms using the connections
introduced in the previous section and the corresponding generalized
Chern-Simons action integrals are discussed. \ In the sixth section it is
shown how a particular class of these connections can be used to construct
action integrals for Einstein's vacuum field equations in four dimensions with
or without a cosmological constant. \ There is a large body of research
devoted to the study of Chern-Simons gravity and related topics with two
papers dating from the 1980's, \cite{deser} and \cite{witten1} being
particularly influential. \ While many of these investigations deal with
gravity in 2+1 dimensions, and are reviewed in \cite{carlip}, results related
to gravity in higher dimensions have also been obtained; some are discussed in
\cite{zanelli} and \cite{jackiw}. \ Many further references can be found in
the last three papers. \ Sections five and six differ from that body of work
in that here use is made of generalized characteristic classes to construct
action integrals for gravitational and other physical systems, an approach
initiated in \cite{tung1} and \cite{tung2} and developed in \cite{rob5} and
\cite{rob6}.

Only type $N=1$ forms are considered here but extensions, where appropriate,
to forms of type $N>1$ is straightforward. \ Discussions of the algebra and
calculus of $N>1$ forms can be found in \cite{rob3}- \cite{rob7} with a type
$N=2$ generalized connection which is flat when Einstein's equations are
satisfied being presented in \cite{rob4}.

The manifolds and geometrical objects considered may be real or complex but in
this paper it will be assumed that the geometry is real, all geometrical
objects are smooth and $M$ is an $n-$dimensional real, smooth, orientable and
oriented manifold. \ Bold-face Roman letters are used to denote generalized
forms and generalized form-valued vector fields,\ ordinary forms on $M$ are
usually denoted by Greek letters and ordinary vector fields on $M$ by lower
case Roman letters. \ Occasionally the degree of a form is indicated above it.
\ The exterior product of any two forms, for example $\alpha$ and $\beta,$ is
written $\alpha\beta$, and as usual, any ordinary $p-$form $\overset{p}{\alpha
}$, with $p$ either negative or greater than $n$, is zero. \ The Einstein
summation convention is used.

\section{Algebra and calculus of type N=1 forms and vector fields}

The algebra and calculus of generalized forms used in this paper are reviewed
in this section using the notation of \cite{rob5} and \cite{rob6} \ A basis
for type $N=1$ generalized forms consists of any basis for ordinary forms on
$M$ augmented by a linearly independent minus one-form $\mathbf{m}$. \ Minus
one forms have the algebraic properties of ordinary exterior forms but are
assigned a degree of minus one. \ They satisfy the ordinary distributive and
associative laws of exterior algebra and the exterior product rule%
\begin{equation}
\overset{p}{\rho}\mathbf{m}=(-1)^{p}\mathbf{m}\overset{p}{\rho};\text{
}\mathbf{m}^{2}=0,\text{ }%
\end{equation}
together with the condition of linear independence. \ Thus, for a given choice
of $\mathbf{m}$, a type $N=1$ generalized p-form, $\overset{p}{\mathbf{r}}%
\in\Lambda_{(1)}^{p}$, can be written as%
\begin{equation}
\mathbf{r}=\rho+\lambda\mathbf{m},
\end{equation}
where $\rho,$ and $\lambda$ are, respectively, ordinary $p-$ and $(p+1)-$forms
and $p$ can take integer values from $-1$ to $n$.

If $\varphi$ is a smooth map between manifolds $P$ and $M,$ $\varphi
:P\rightarrow M,$ then the induced map of type $N=1$ generalized forms,
$\varphi_{(1)}^{\ast}:\Lambda_{(1)}^{p}(M)\rightarrow\Lambda_{(1)}^{p}(P)$, is
the linear map defined by using the standard pull-back map, $\varphi^{\ast}$,
for ordinary forms%
\begin{equation}
\varphi_{(1)}^{\ast}(\mathbf{r})=\varphi^{\ast}(\rho)+\varphi^{\ast}%
(\lambda)\mathbf{m},
\end{equation}
and $\varphi_{(1)}^{\ast}(\overset{p}{\mathbf{r}}\overset{q}{\mathbf{s}%
})=\varphi_{(1)}^{\ast}(\overset{p}{\mathbf{r}})\varphi_{(1)}^{\ast
}(\overset{q}{\mathbf{s}})$. \ Hence $\varphi_{(1)}^{\ast}(\mathbf{m)=m}$.

Henceforth in this paper, in addition to assuming that the exterior derivative
of generalized forms satisfies the usual properties, it is\ assumed, without
loss of generality, \cite{rob8}, that%
\begin{equation}
d\mathbf{m}=\epsilon,
\end{equation}
where $\epsilon$ denotes a real constant$.$

The exterior derivatives of a type $N=1$ generalized form
$\overset{p}{\mathbf{r}}$ is then%
\begin{equation}
d\mathbf{r}=[d\rho+(-1)^{p+1}\epsilon\lambda]+d\lambda\mathbf{m},
\end{equation}
where $d$ is the ordinary exterior derivative when acting on ordinary forms.
\ The exterior derivative $d:$ $\Lambda_{(1)}^{p}(M)\rightarrow\Lambda
_{(1)}^{p+1}(M)$ is an anti-derivation of degree one,%
\begin{align}
d(\overset{p}{\mathbf{r}}\overset{q}{\mathbf{s}})  &
=(d\overset{p}{\mathbf{r}})\overset{q}{\mathbf{s}}+(-1)^{p}%
\overset{p}{\mathbf{r}}d\overset{q}{\mathbf{s}},\\
d^{2}  &  =0.\nonumber
\end{align}
and $(\Lambda_{(N)}^{\bullet}(M),d)$ is a differential graded algebra.

The dual of a generalized one-form is a generalized form-valued vector field,
\cite{rob8}. \ Such a type $N=1$ vector field is defined on a coordinate patch
$U$ $\subseteq M$ by%
\begin{equation}
\mathbf{V=v}^{\rho}\frac{\partial}{\partial x^{\rho}}=(v^{\rho}+v_{\sigma
}^{\rho}dx^{\sigma}\mathbf{m)}\frac{\partial}{\partial x^{\rho}}=v+(v_{\sigma
}^{\rho}dx^{\sigma}\mathbf{m)}\frac{\partial}{\partial x^{\rho}},
\end{equation}
where $\{x^{\alpha}\}$ are local coordinates on $U$ and $v=v^{\rho}%
\frac{\partial}{\partial x^{\rho}}$ is an ordinary vector field. \ Globally
$\mathbf{V}$ is determined by an ordinary vector field $v$ and a $(1,1)$ type
tensor field given in local coordinates by $v_{\sigma}^{\rho}\frac{\partial
}{\partial x^{\rho}}\otimes dx^{\sigma}$ on $M$. \ The set of all such vector
fields in $M$, $\{\mathbf{V\}}$, is naturally a module, $\mathcal{V}_{(1)}%
(M)$, over the generalized zero forms on $M$, $\Lambda_{(1)}^{0}(M)$.

\textbf{Example 1: \ }In Euclidean three-space, with Euclidean coordiates and
metric $ds^{2}=\delta_{\mu\nu}dx^{\mu}dx^{\nu}$ the scalar or dot product of
two generalized form-valued vector fields $\mathbf{V=v}^{\rho}\frac{\partial
}{\partial x^{\rho}}$ and $\mathbf{W}=\mathbf{w}^{\rho}\frac{\partial
}{\partial x^{\rho}}=(w^{\rho}+w_{\sigma}^{\rho}dx^{\sigma}\mathbf{m)}%
\frac{\partial}{\partial x^{\rho}}=w+(w_{\sigma}^{\rho}dx^{\sigma}%
\mathbf{m)}\frac{\partial}{\partial x^{\rho}}$\ is the generalized zero form
\[
\mathbf{V.W}=\delta_{\mu\nu}\mathbf{v}^{\mu}\mathbf{w}^{\nu}=v.w+(\delta
_{\mu\nu}v^{\mu}w_{\sigma}^{\nu}+\delta_{\mu\nu}w^{\mu}v_{\sigma}^{\nu
})dx^{\sigma}\mathbf{m}%
\]
The vector or crossproduct is the generalized form-valued vector field
\[
\mathbf{V\times W=}\varepsilon^{\rho\mu\nu}\mathbf{v}^{\mu}\mathbf{w}^{\nu
}\frac{\partial}{\partial x^{\rho}}=v\times w+[\varepsilon^{\rho\mu\nu}%
(v^{\mu}w_{\sigma}^{\nu}+w^{\nu}v_{\sigma}^{\mu})dx^{\sigma}\mathbf{m]}%
\frac{\partial}{\partial x^{\rho}}%
\]
where $\varepsilon^{\rho\mu\nu}$ is the totally skew-symmetric Levi-Civita symbol.

The interior product of a generalized $p-$form $\mathbf{r}=\rho+\lambda
\mathbf{m}$ with respect to $\mathbf{V}$ is given by the formula%
\begin{equation}
i_{\mathbf{V}}\mathbf{r}=\mathbf{v}^{\alpha}i_{\frac{\partial}{\partial
x^{\alpha}}}\mathbf{r.}%
\end{equation}
where $i_{_{\frac{\partial}{\partial x^{\alpha}}}}\mathbf{r}$ is the interior
product of $\mathbf{r}$ by the ordinary vector field $\frac{\partial}{\partial
x^{\alpha}}$ as defined in \cite{rob2}, that is $i_{\frac{\partial}{\partial
x^{\alpha}}}\mathbf{r}=i_{\frac{\partial}{\partial x^{\alpha}}}\rho
+i_{\frac{\partial}{\partial x^{\alpha}}}\lambda\mathbf{m}$. \ For $p$ equal
to minus one and zero
\begin{align}
i_{\mathbf{V}}\overset{-1}{\mathbf{r}}  &  =0.\\
i_{\mathbf{V}}\overset{0}{\mathbf{r}}  &  =\lambda_{\alpha}v^{\alpha
}\mathbf{m},\nonumber
\end{align}
and for $p\geqq1$%
\begin{align}
i_{\mathbf{V}}\mathbf{r}  &  =i_{v}\mathbf{r+}\overset{p}{\gamma}%
\mathbf{m}=i_{v}\rho+i_{v}\lambda\mathbf{m+}\overset{p}{\gamma}\mathbf{m,}\\
\overset{p}{\gamma}  &  =(-1)^{p-1}v_{\beta}^{\alpha}dx^{\beta}(i_{\frac
{\partial}{\partial x^{\alpha}}}\rho)\nonumber\\
&  =\frac{(-1)^{p-1}}{(p-1)!}v_{\lambda_{1}}^{\alpha}\rho_{\alpha\lambda
_{2}....\lambda_{p}}dx^{\lambda_{1}...}dx^{\lambda_{p}}.\nonumber
\end{align}
This interior product satisfies the graded Leibniz rule%
\begin{equation}
i_{\mathbf{V}}(\overset{p}{\mathbf{r}}\overset{q}{\mathbf{s}})=(i_{\mathbf{V}%
}\overset{p}{\mathbf{r}})\overset{q}{\mathbf{s}}+(-1)^{p}%
\overset{p}{\mathbf{r}}(i_{\mathbf{V}}\overset{q}{\mathbf{s}}),
\end{equation}
but does not in general anti-commute because the interior product on
generalized zero forms need not be zero,%
\begin{equation}
(i_{\mathbf{W}}\mathbf{\circ}i_{\mathbf{V}}+i_{\mathbf{V}}\mathbf{\circ
}i_{\mathbf{W}})\mathbf{r}\mathbf{=}(-1)^{p-1}\{[v_{\beta}^{\alpha}w^{\beta
}+w_{\beta}^{\alpha}v^{\beta}](i_{\frac{\partial}{\partial x^{\alpha}}}%
\rho\}\mathbf{m,}%
\end{equation}
where $\mathbf{W}=(w^{\rho}+w_{\sigma}^{\rho}dx^{\sigma}\mathbf{m)}%
\frac{\partial}{\partial x^{\rho}}$. \ The Lie derivative of generalized forms
by generalized form-valued vector fields is a derivation of degree zero
defined by $\pounds _{\mathbf{V}}=d\circ i_{\mathbf{V}}+i_{\mathbf{V}}\circ d$
\ and the Lie bracket of two generalized form-valued vector fields
$\mathbf{V}$ and $\mathbf{W}$ on $M$ is the generalized form-valued vector
field, $[\mathbf{V,W}]$, defined by the relation $(\pounds _{\mathbf{V}%
}\pounds _{\mathbf{W}}-\pounds _{\mathbf{W}}\mathbf{\pounds _{\mathbf{V}%
})r=\pounds }_{[\mathbf{V,W]}}\mathbf{r}$. \ Type $N=1$ generalized
form-valued vector fields include, as special cases, both ordinary vector
fields, \cite{rob2}, and the previously introduced generalized vector fields,
\cite{lah1} and \cite{chat}.

Just as the algebra and differential calculus of ordinary differential forms
on $M$ can be expressed in terms of functions and vector fields on the reverse
parity tangent bundle, $\Pi TM$, of $M$, \cite{voronov} and \cite{witten3}
generalized forms can be represented in terms of functions and vector fields
on the Whitney sum of $\Pi TM$ and a trivial reverse parity line bundle over
$M$, that is\ a trivial line bundle with fibre $\mathbb{R}^{1}$ replaced by
$\mathbb{R}^{0\mid1}$. \ \ Further details about this and type $N$ generalized
form-valued vector fields are in \cite{rob8}.

Integration is defined using polychains, \cite{rob6}. \ A $p-$polychain of
type $N=1$ in $M$, written here as $\mathbf{c}_{p}$, is an ordered pair of
ordinary (real, singular) chains in $M$%
\begin{equation}
\mathbf{c}_{p}=(c_{p},c_{p+1}),
\end{equation}
where $c_{p}$ is an ordinary $p-$chain, and $c_{p+1}$ is an ordinary
$p+1-$chains. \ The ordinary chains have respective boundaries $\partial
c_{p},\partial c_{p+1}$, and the boundary of the polychain $\mathbf{c}_{p}$ is
the $(p-1)-$polychain%
\begin{equation}
\partial\mathbf{c}_{p}=(\partial c_{p},\partial c_{p+1}+(-1)^{p}\epsilon
c_{p}),
\end{equation}
and%
\begin{equation}
\partial^{2}\mathbf{c}_{p}=0.
\end{equation}
\textbf{Example 2: }When a polychain is determined by just one ordinary chain
as in the three examples%
\begin{align*}
\mathbf{c}_{p}  &  =(0,c_{p+1}),\\
\mathbf{c}_{p}  &  =(c_{p},0),\\
\mathbf{c}_{p}  &  =(\pm\partial c_{p+1},c_{p+1}),
\end{align*}
then the corresponding three boundaries are%
\begin{align*}
\partial\mathbf{c}_{p}  &  =(0,\partial c_{p+1}),\\
\partial\mathbf{c}_{p}  &  =(\partial c_{p},(-1)^{p}\epsilon c_{p}),\\
\partial\mathbf{c}_{p}  &  =(0,[1\pm(-1)^{p}\epsilon]\partial c_{p}).
\end{align*}

When $N=1$ the integral of a generalized form $\overset{p}{\mathbf{a}}$ over a
polychain $\mathbf{c}_{p}$ is%
\begin{equation}
\int_{\mathbf{c}_{p}}\overset{p}{\mathbf{r}}=\int_{c_{p}}\rho+\int_{c_{p+1}%
}\lambda.
\end{equation}
Stokes' theorem for generalized forms and polychains in the type $N=1$ case
states that%
\begin{equation}
\int_{\mathbf{c}_{p}}d\overset{p-1}{\mathbf{r}}=\int_{\partial\mathbf{c}_{p}%
}\overset{p-1}{\mathbf{r}}.
\end{equation}

\section{Generalized connections}

Generalized connection with values in the Lie algebra of a matrix Lie group
$G$ are defined in essentially the same way as ordinary connections
\cite{nak}, except that ordinary forms, including zero forms, are replaced by
generalized forms. \ Let $P$ be a principal $G$ bundle over $M$. \ If
$\{U_{I}\}$ is a covering of an $n-$dimensional manifold $M$ by coordinate
charts, then a generalized connection $\mathbf{A}$ with values in the Lie
algebra $\mathfrak{g}$ of the matrix Lie group $G$ is an assignment of a
$\mathfrak{g}$-valued generalized one-form, $\mathbf{A}_{I}$, to each set
$U_{I}$ and such that on $U_{I}\cap U_{J}$ , for all $I$ and $J$,
\begin{equation}
\mathbf{A}_{J}=(t_{IJ}^{-1})dt_{IJ}+(t_{IJ}^{-1})\mathbf{A}_{I}t_{IJ},
\end{equation}
where $t_{IJ}:U_{I}\cap U_{J}\rightarrow G$,\ by $p\rightarrow t_{IJ}(p)$ are
transition functions satisfying the usual conditions, $t_{II}(p)=1$, $p\in
U_{I},$ $t_{IJ}(p)=[t_{JI}(p)]^{-1}$ $p\in U_{I}\cap U_{J}$, $t_{IJ}%
(p)t_{JK}(p)=t_{IK}(p)$ $p\in U_{I}\cap U_{J}\cap U_{K}$. \ Transition
functions $\{t_{IJ}\}$ and $\{\widetilde{t}_{IJ}\}$ are (gauge) equivalent
when $\widetilde{t}_{IJ}=(g_{I})^{-1}t_{IJ}g_{J}$ and $g_{I}:U_{I}\rightarrow
G$ and $g_{J}$ $:U_{J}\rightarrow G$ determine gauge transformations,
$\mathbf{A}_{I}\rightarrow(g_{I}^{-1})dg_{I}+(g_{I}^{-1})\mathbf{A}_{I}g_{I}$
and $\mathbf{A}_{J}\rightarrow(g_{J}^{-1})dg_{J}+(g_{J}^{-1})\mathbf{A}%
_{J}g_{J}$ in $U_{I}$ and $U_{J}$ respectively.

The curvature two-form $\mathbf{F}_{I}$ is the generalized form%
\begin{equation}
\mathbf{F}_{I}\mathbf{=dA}_{I}+\mathbf{A}_{I}\mathbf{A}_{I},
\end{equation}
and under the transformation in Eq.(18)%
\begin{equation}
\mathbf{F}_{J}\mathbf{=}(t_{IJ}^{-1})\mathbf{F}_{I}t_{IJ}.
\end{equation}
On any coordinate chart such as $U_{I}$ the connection one-form $\mathbf{A}%
_{I}$ can be written as \
\begin{equation}
\mathbf{A}_{I}=\alpha_{I}+\beta_{I}\mathbf{m,}%
\end{equation}
where $\alpha_{I}$ and $\beta_{I}$ are respectively ordinary matrix valued
one-forms and two-forms on $M$ and the curvature two-form is
\begin{align}
\mathbf{F}_{I}  &  =\mathcal{F}_{I}+\epsilon\beta_{I}+D\beta_{I}\mathbf{m,}\\
\mathcal{F}_{I}  &  =d\alpha_{I}+\alpha_{I}\alpha_{I},\nonumber\\
D\beta_{I}  &  =d\beta_{I}+\alpha_{I}\beta_{I}-\beta_{I}\alpha_{I}.\nonumber
\end{align}
It follows from the above that the locally defined ordinary one-forms
$\alpha_{I}$ and curvature two-forms $\mathcal{F}_{I},$ patch together to
define global connection and curvature forms, $\alpha$ and $\mathcal{F}$ , of
an ordinary connection. \ The ordinary two-forms $\beta$ transform as
Lie-algebra valued two-forms.

These ideas can be be broadened by extending the structure group. \ Let $G$ be
a matrix Lie group with Lie algebra $\mathfrak{g}$ as above. \ Let
$G_{(0)}=\{g\}$ be the space of $G$-valued (ordinary) zero-forms belonging to
$\Lambda_{(0)}^{0}(U)$ where $U$ is an open set $U\subseteq M$. \ This is a
group under multiplication with identity the unit matrix $1$. \ Let
$\mathfrak{g}_{(0)\text{ }}=\{h\}$ be the set of $\mathfrak{g}$-valued
one-forms $\in$ $\Lambda_{(0)}^{1}(U)$. \ There is an ad-action of of
$G_{(0)}$ on $\mathfrak{g}_{(0)},$ that is a homomorphism $\Phi:$
$G_{(0)}\rightarrow aut(\mathfrak{g}_{(o)})$, $\Phi($ $g):h\longrightarrow
gh(g)^{-1}.$ \ Then the set of type $N=1$ matrix-valued generalized
zero-forms, $G_{(1)}=\{\mathbf{g\mid g}=\mathbf{h}g=(1+h\mathbf{m})g\}$ is a
group under matrix and exterior multiplication. \ Calculation shows that the
product of $\mathbf{g}_{1}=\mathbf{h}_{1}g_{1}=(1+h_{1}\mathbf{m})g_{1}$ and
$\mathbf{g}_{2}=\mathbf{h}_{2}g_{2}=(1+h_{2}\mathbf{m})g_{2}\in G_{(1)}$ is
the element of $G_{(1)}$ given by \
\begin{equation}
\mathbf{g}_{1}\mathbf{g}_{2}=\{1+[h_{1}+g_{1}h_{2}(g_{1})^{-1}]\mathbf{m\}}%
g_{1}g_{2}.
\end{equation}
The identity of $G_{(1)}$ is the unit matrix $1,$ and the inverse of
$\mathbf{g}$ is given by
\begin{equation}
(\mathbf{g})^{-1}=\{1-[(g)^{-1}hg]\mathbf{m}\}(g)^{-1}.
\end{equation}
The Lie algebra $\mathfrak{g}_{(1)}$ of $G_{(1)}$ is given by $\{l\mid
l=\lambda+\mu\mathbf{m\}}$ where $\lambda$ and $\mu$ are respectively ordinary
zero and one forms taking values in $\mathfrak{g}$.

Under a generalized gauge transformation by $\mathbf{g}$ where $\mathbf{g}\in$
$G_{(1)}$%
\begin{align}
\mathbf{A}  &  \rightarrow(\mathbf{g}^{-1})d\mathbf{g}+(\mathbf{g}%
^{-1})\mathbf{Ag=}=g^{-1}dg+g^{-1}[\mathbf{h}^{-1}d\mathbf{h+h}^{-1}%
\mathbf{Ah}]g\\
&  =g^{-1}dg+g^{-1}(\alpha-\epsilon h)g+g^{-1}[dh-\epsilon hh+h\alpha+\alpha
h+\beta]g\mathbf{m}\nonumber\\
\mathbf{F}  &  \rightarrow(\mathbf{g}^{-1})\mathbf{Fg}=g^{-1}\{\mathcal{F+}%
\epsilon\mathcal{\beta+[}D\beta+(\mathcal{F+}\epsilon\mathcal{\beta
)}h\mathcal{-}h\mathcal{(F+}\epsilon\mathcal{\beta)]}\mathbf{m\}}g.\nonumber
\end{align}
The definition of a global generalized connection given above may be
generalized in the obvious way. \ Briefly, for a covering $\{U_{I}\}$ of $M$
by coordinate charts, specify transition functions $\mathbf{t}_{IJ}\in
G_{(1)}$ on $U_{I}\cap U_{J}$ by $p\rightarrow\overset{0}{\mathbf{t}}_{IJ}(p)$
satisfying $\overset{0}{\mathbf{t}}_{II}(p)=1$, $p\in U_{I},$
$\overset{0}{\mathbf{t}}_{IJ}(p)=[\overset{0}{\mathbf{t}}_{JI}(p)]^{-1}$ $p\in
U_{I}\cap U_{J}$, $\overset{0}{\mathbf{t}}_{IJ}(p)\overset{0}{\mathbf{t}}%
_{JK}(p)=\overset{0}{\mathbf{t}}_{IK}(p)$ $p\in U_{I}\cap U_{J}\cap U_{K}$ and
local generalized connection one-forms $\mathbf{A}_{I}$, on each $U_{I}$,
related by $\mathbf{A}_{J}=(\mathbf{t}_{IJ})^{-1}\mathbf{A}_{I}\mathbf{t}%
_{IJ}+(\mathbf{t}_{IJ})^{-1}d\mathbf{t}_{IJ}$ on $U_{I}\cap U_{J}$.
\ Transition functions $\{\mathbf{t}_{IJ}\}$ and $\{\widetilde{\mathbf{t}%
}_{IJ}\}$ are (gauge) equivalent when $\widetilde{\mathbf{t}}_{IJ}%
=(\mathbf{g}_{I})^{-1}\mathbf{t}_{IJ}\mathbf{g}_{J}$ and $\mathbf{g}_{I}$ and
$\mathbf{g}_{J}$ respectively determine generalized gauge transformations in
$U_{I}$ and $U_{J}$ as in Eq.(25) above.

Henceforth connections on $M$ will be discussed and the subscripts
corresponding to coordinate charts will be dropped.

The curvature satisfies the Bianchi identities%
\begin{equation}
\mathbf{D\mathbf{F}=}d\mathbf{F}+\mathbf{AF}-\mathbf{FA}=0,
\end{equation}
\ where here $\mathbf{D}$ denotes the covariant exterior derivative of a type
$N=1$ valued generalized form. \ For a $\mathfrak{g}-$valued p-form
$\mathbf{P}$
\begin{equation}
\mathbf{DP=}d\mathbf{P}+\mathbf{AP}+(-1)^{p+1}\mathbf{PA}.
\end{equation}
Let $E$ be a $d-$dimensional vector bundle associated to $P$\ and let
$\{e_{i}\}$, $i=1$ to $d$ be a a basis of sections of $E$ over $U\subseteq M$
with the usual action of $g\in$ $G$, $e_{i}\rightarrow e_{j}g_{i}^{j}$.  A
type $N=1$ generalized form-valued vector field is given by $\mathbf{V=v}%
^{i}e_{i}=(v^{i}+v_{\sigma}^{i}dx^{\sigma}\mathbf{m)}e_{i}$, where the
components $\mathbf{v}^{i}$ are generalized zero forms on $U$, and the action
of $\mathbf{g}\in$ $G_{(1)}$ is the extension of the action of $g\in$ $G$
given by $e_{i}\rightarrow e_{j}\mathbf{g}_{i}^{j}$ , $\mathbf{v}%
^{i}\rightarrow(\mathbf{g}^{-1})_{j}^{i}\mathbf{v}^{j}$. \ The covariant
derivative of $\mathbf{V}$ is given by%
\begin{equation}
\mathbf{DV}=\mathbf{Dv}^{i}\mathbf{\otimes}e_{i}=(d\mathbf{v}^{i}%
+\mathbf{\mathbf{A}}_{j}^{i}\mathbf{v}^{j}\mathbf{)\otimes}e_{i}.
\end{equation}
In local coordinates $\{x^{\nu}\}$ on $M$,%

\begin{equation}
\mathbf{Dv}^{i}\mathbf{=}Dv^{i}-\epsilon v_{\nu}^{i}dx^{\nu}+[D(v_{\nu}%
^{i}dx^{\nu})+\beta_{j}^{i}v^{j}]\mathbf{m,}%
\end{equation}
and $\mathbf{D}$ and $D$ are the covariant exterior derivatives with respect
to $\mathbf{A}$ and $\alpha$ (with matrix representations with $(i,j)$ entries
$\mathbf{A}_{j}^{i}$ and $\alpha_{j\text{ }}^{i}$respectively). \ The
covariant derivative with respect to a type $N=1$ generalized form-valued
vector field $\mathbf{W}$, is the generalized form-valued vector field\textsf{
}%
\begin{equation}
\mathbf{D}_{\mathbf{W}}\mathbf{V=[}i_{\mathbf{W}}(d\mathbf{v}^{i}%
+\mathbf{A}_{j}^{i}\mathbf{v}^{i}\mathbf{)]}e_{i}.
\end{equation}
The covariant derivative is extended to generalized form -valued tensor fields
by using the linearity and product rules satisfied by ordinary covariant
derivatives and tensor fields.

A field $\mathbf{V}$ is a parallel vector field if $\mathbf{DV}=0$, that is%

\begin{align}
Dv^{i}-\epsilon v_{\nu}^{i}dx^{\nu}  &  =0,\\
D(v_{\nu}^{i}dx^{\nu})+\beta_{j}^{i}v^{j}  &  =0.\nonumber
\end{align}
and such a system of equations is completely integrable if and only if the
generalized curvature $\mathbf{\mathbf{F}}$ is zero, that is when%
\begin{align}
\mathcal{F}_{j}^{i}\mathcal{+}\epsilon\beta_{j}^{i}  &  \mathcal{=}0,\\
D\beta_{j}^{i}  &  =0.\nonumber
\end{align}

The notion of parallel transport on a sub-manifold is defined by considering
the pull-backs of these equations to the sub-manifold.

When $\mathbf{F=0}$ the connection $\mathbf{A=}$ $(\mathbf{g}_{c}%
\mathbf{g})^{-1}d(\mathbf{g}_{c}\mathbf{g)}$ for some $\mathbf{g}$ and any
closed $\mathbf{g}_{c}\mathbf{\in}G_{(1)}$, \cite{rob5}. \ That is if
$\mathbf{g}=(1+h\mathbf{m})g$ and $\mathbf{g}_{c}=(1+h_{c}\mathbf{m})g_{c}$
where $dg_{c}=\epsilon h_{c}g_{c}$ and $d(h_{c}g_{c})=0,$%
\begin{equation}
\mathbf{A}=(\mathbf{g}_{c}\mathbf{g})^{-1}d(\mathbf{g}_{c}\mathbf{g)=}%
(g)^{-1}dg-\epsilon(g)^{-1}hg+\{(g)^{-1}[dh-\epsilon hh]g\}\mathbf{m,}%
\end{equation}
and any parallel vector field $\mathbf{V}=(\mathbf{g}^{-1})_{j}^{i}%
\mathbf{v}_{0}^{j}e_{i}$ for some closed generalized zero-forms $\mathbf{v}%
_{0}^{j}$.

Henceforth in this paper only groups such that the connections $\mathbf{A}$
have, $Tr\mathbf{A}$, zero will be considered.

The generalized Chern-Pontrjagin class is determined by a generalized
four-form $\mathbf{CP}$%
\begin{equation}
\mathbf{CP}=\frac{1}{8\pi^{2}}Tr(\mathbf{FF}),
\end{equation}
\ which is equal to the exterior derivative of the generalized Chern-Simons
three-form $\mathbf{CS}$ where%
\begin{equation}
\mathbf{CS}=\frac{1}{8\pi^{2}}Tr(\mathbf{AF}-\frac{1}{3}\mathbf{AAA}).
\end{equation}
More generally, if $\mathbf{k}$ is any closed generalized zero-form, then%
\begin{equation}
d(\mathbf{kCS})=\mathbf{kCP}\text{.}%
\end{equation}
By Stokes' theorem, Eq.(17),\ for a polychain $\mathbf{c}_{4}$%
\begin{equation}
\int_{\mathbf{c}_{4}}\mathbf{kCP}=\int_{\mathbf{\partial c}_{4}}\mathbf{kCS}.
\end{equation}
Under the generalized gauge transformation given by Eq.(25)%
\begin{align}
\mathbf{CP}  &  \mathbf{\rightarrow CP,}\\
\mathbf{CS}  &  \mathbf{\rightarrow CS-}\frac{1}{8\pi^{2}}d\{Tr[(d\mathbf{g}%
)(\mathbf{g})^{-1}\mathbf{A]\}-}\frac{1}{24\pi^{2}}Tr[(\mathbf{g}%
^{-1}d\mathbf{g})^{3}].\nonumber
\end{align}
Since the last (generalized winding number) term is closed when $\mathbf{c}%
_{3}=\partial\mathbf{c}_{4}$%
\begin{equation}
\int_{\mathbf{c}_{3}}\mathbf{CS\rightarrow}\int_{\mathbf{c}_{3}}\mathbf{CS}.
\end{equation}

For $\mathbf{A}=\alpha+\beta\mathbf{m}$ as above%
\begin{equation}
\mathbf{CS}\mathbf{=}\frac{1}{8\pi^{2}}Tr[(\alpha\mathcal{F}-\frac{1}{3}%
\alpha\alpha\alpha+\epsilon\alpha\beta)+(\alpha D\beta+\beta\mathcal{F}%
-\beta\alpha\alpha+\epsilon\beta\beta)\mathbf{m]}.
\end{equation}
In sections five and six type $N=1$ Chern-Simons integrals for a polychain
\begin{equation}
\mathbf{c}_{3}=\partial\mathbf{c}_{4}=\partial(c_{4},c_{5})=(\partial
c_{4},\partial c_{5}+\epsilon c_{4}),
\end{equation}
will be used as action integrals. \ In this case, \cite{rob6}, $\int%
_{\mathbf{c}_{4}}\mathbf{CP}$ $=\mathbf{=}\int_{\partial\mathbf{c}_{4}%
}\mathbf{CS=}\int_{\mathbf{c}_{3}}\mathbf{CS}$ and%
\begin{equation}
\int_{\mathbf{c}_{3}}\mathbf{CS}=\frac{1}{8\pi^{2}}[\int_{\partial c_{4}%
}Tr(\alpha\mathcal{F}-\frac{1}{3}\alpha\alpha\alpha)+\int_{\partial
c_{5}+\epsilon c_{4}}Tr(2\beta\mathcal{F}+\epsilon\beta\beta)\mathbf{].}%
\end{equation}

The variation of $\mathbf{A}=\alpha+\beta\mathbf{m}$\ is $\delta$
$\mathbf{A}=\delta\alpha+\delta\beta\mathbf{m}$. \ Then from Eqs.(35) and (42)%
\begin{align}
\delta\mathbf{CS}  &  =\frac{1}{8\pi^{2}}[Tr(2\delta\mathbf{AF)}%
+d(Tr\delta\mathbf{AA})],\\
\delta\int_{\mathbf{c}_{3}}\mathbf{CS}  &  \mathbf{=}\frac{1}{8\pi^{2}}%
\{\int_{\partial c_{4}}Tr[2\delta\alpha(\mathcal{F+\epsilon\beta)]+}%
\int_{\partial c_{5}+\epsilon c_{4}}2Tr[\delta\alpha D\beta+\delta
\beta(\mathcal{F+\epsilon\beta)]}\}.\nonumber
\end{align}

\section{Connections, metrics and gravity}

Consider, on an $n$ dimensional manifold $M$, type $N=1$ generalized
connections represented by $(p+q+1,p+q+1)$ matrix valued generalized one-forms%
\begin{equation}
\mathbf{A=}\left(
\begin{array}
[c]{cc}%
\mathbf{A}_{b}^{a} & -\sigma\mathbf{A}^{a}\\
\mathbf{A}_{b} & 0
\end{array}
\right)  ,
\end{equation}
where $\sigma$ is either $1$, $-1$ or $0$ and $\mathbf{A}$ takes values in the
Lie algebra of $G=SO(p+1,q)$ when $\sigma=1$, $G=SO(p,q+1)$ when $\sigma=-1$
and $G=ISO(p,q)$ when $\sigma=0$. \ In the first two cases the metric is given
by the $(p+q+1)\times(p+q+1)$ matrix%
\begin{equation}
\left(
\begin{array}
[c]{cc}%
\eta_{ab} & 0\\
0 & \sigma
\end{array}
\right)  ,
\end{equation}%
\[
(\eta_{ab})=\left(
\begin{array}
[c]{cc}%
1_{p\times p} & 0\\
0 & -1_{q\times q}%
\end{array}
\right)  ,
\]
and $\mathbf{A}_{b}=\eta_{ba}\mathbf{A}^{a}$. \ Latin indices ranging and
summing from $1$ to $p+q$. \ The curvature of $\mathbf{A}$ is given by%
\begin{equation}
\mathbf{F}=d\mathbf{A}+\mathbf{AA}=\left(
\begin{array}
[c]{cc}%
\mathbf{F}_{b}^{a} & -\sigma\mathbf{F}^{a}\\
\mathbf{F}_{b} & 0
\end{array}
\right)  ,
\end{equation}
where%
\begin{align}
\mathbf{F}_{b}^{a}  &  =\mathbf{dA}_{b}^{a}+\mathbf{A}_{c}^{a}\mathbf{A}%
_{b}^{c}-\sigma\mathbf{A}^{a}\mathbf{A}_{b},\\
\mathbf{F}^{a}  &  =d\mathbf{A}^{a}+\mathbf{A}_{b}^{a}\mathbf{A}^{b},\text{
}\mathbf{F}_{b}=\eta_{bc}\mathbf{F}^{c}.\nonumber
\end{align}

Now let%
\begin{align}
\mathbf{A}_{b}^{a}  &  =\omega_{b}^{a}-\kappa_{b}^{a}\mathbf{m,}\\
\mathbf{A}^{a}  &  =\frac{1}{l}(\theta^{a}-\Theta^{a}\mathbf{m),}\nonumber
\end{align}
where $l$ is a non-zero constant and $\omega_{ab}=-\omega_{ba}$, $\kappa
_{ab}=-\kappa_{ba}$ are, respectively, an ordinary $so(p,q)$-valued one-form
and two-form and $\theta^{a}$ and $\Theta^{a}$ are, respectively, ordinary
one-forms and two-forms on $M$. \ Then%
\begin{align}
\mathbf{F}_{b}^{a}  &  =\Omega_{b}^{a}-\frac{\sigma}{l^{2}}\theta^{a}%
\theta_{b}-\epsilon\kappa_{b}^{a}-[D\kappa_{b}^{a}+\frac{\sigma}{l^{2}}%
(\Theta^{a}\theta_{b}-\theta^{a}\Theta_{b})]\mathbf{m,}\\
\mathbf{F}^{a}  &  =\frac{1}{l}[D\theta^{a}-\epsilon\Theta^{a}+(\kappa_{b}%
^{a}\theta^{b}-D\Theta^{a})\mathbf{m]},\nonumber\\
\Omega_{b}^{a}  &  =d\omega_{b}^{a}+\omega_{c}^{a}\omega_{b}^{c},\nonumber\\
\mathbf{dA}_{b}^{a}+\mathbf{A}_{c}^{a}\mathbf{A}_{b}^{c}  &  =\Omega_{b}%
^{a}-\epsilon\kappa_{b}^{a}-D\kappa_{b}^{a}\mathbf{m,}\nonumber
\end{align}
The covariant exterior derivative with respect to $\omega_{b}^{a}$ is denoted
$D$ so that%
\begin{align*}
D\theta^{a}  &  =d\theta^{a}+\omega_{b}^{a}\theta^{b},D\Theta^{a}=d\Theta
^{a}+\omega_{b}^{a}\Theta^{b},\\
D\kappa_{b}^{a}  &  =d\kappa_{b}^{a}+\omega_{c}^{a}\kappa_{b}^{c}-\kappa
_{c}^{a}\omega_{b}^{c}.
\end{align*}
Under a generalized gauge transformation by $\mathbf{g}=(1+h\mathbf{m}%
\mathbb{)}$, where $h$ is a $\mathfrak{g}$-valued one-form%
\begin{equation}
h=\left(
\begin{array}
[c]{cc}%
h_{b}^{a} & -\sigma\frac{h^{a}}{l}\\
\frac{h_{b}}{l} & 0
\end{array}
\right)  ,\text{ }h_{ab}=-h_{ba},
\end{equation}
the generalized connection gauge transformation, Eq.(25), becomes%
\[
\mathbf{A=}\alpha+\beta\mathbf{m}\rightarrow(\alpha-\epsilon h)+[dh-\epsilon
hh+h\alpha+\alpha h+\beta]\mathbf{m},
\]
that is%
\begin{align}
\omega_{b}^{a}  &  \rightarrow\omega_{b}^{a}-\epsilon h_{b}^{a},\\
\kappa_{b}^{a}  &  \rightarrow\kappa_{b}^{a}-Dh_{b}^{a}+\epsilon h_{c}%
^{a}h_{b}^{c}+\frac{\sigma}{l^{2}}(h^{a}\theta_{b}-h_{b}\theta^{a}-\epsilon
h^{a}h_{b}),\nonumber\\
\theta^{a}  &  \rightarrow\theta^{a}-\epsilon h^{a},\nonumber\\
\Theta^{a}  &  \rightarrow\Theta^{a}-Dh^{a}-\epsilon h^{b}h_{b}^{a}-h_{b}%
^{a}\theta^{b},\nonumber
\end{align}
where
\[
Dh_{b}^{a}=dh_{b}^{a}+h_{c}^{a}\omega_{b}^{c}+\omega_{c}^{a}h_{b}^{c},\text{
}Dh^{a}=dh^{a}+\omega_{b}^{a}h^{b}.
\]
When the one-forms $\{\theta^{a}\}$ constitute a basis in terms of which
$\Theta^{a}=\frac{1}{2}\Theta_{bc}^{a}\theta^{b}\theta^{c}$, where
$\Theta_{bc}^{a}=\Theta_{\lbrack bc]}^{a}$, and $h_{b}^{a}=h_{bc}^{a}%
\theta^{c}$, where $h_{abc}=h_{[ab]c}$, it is straighforward to show that
under the generalized gauge transformation given by Eq.(51) with%
\begin{align}
h_{abc}  &  =\frac{1}{2}(\Theta_{cab}+\Theta_{bac}+\Theta_{acb})\\
h^{a}  &  =0,\nonumber
\end{align}
the two-forms $\Theta^{a}$ transform to zero, $\Theta^{a}\rightarrow0$.

The generalized gauge condition $\Theta^{a}=0$ is preserved by gauge
transformations given by $g\in G$ where%
\begin{equation}
g=\left(
\begin{array}
[c]{cc}%
g_{b}^{a} & 0\\
0 & 1
\end{array}
\right)  ,\text{ }g_{b}^{a}g_{d}^{c}\eta_{ac}=\eta_{bd}%
\end{equation}
if $\sigma$ is non-zero and by $g\in ISO(p,q)$ if $\sigma=0$.

The generalized connection $\mathbf{A}$ is flat if and only if $\mathbf{F}=0$
and then%
\begin{align}
D\theta^{a}  &  =\epsilon\Theta^{a},\\
\Omega_{b}^{a}  &  =\frac{\sigma}{l^{2}}\theta^{a}\theta_{b}+\epsilon
\kappa_{b}^{a},\nonumber\\
D\Theta^{a}  &  =\kappa_{b}^{a}\theta^{b},\nonumber\\
D\kappa_{b}^{a}  &  =\frac{\sigma}{l^{2}}(\theta^{a}\Theta_{b}-\Theta
^{a}\theta_{b}).\nonumber
\end{align}
This is a closed differential ideal.

Suppose now that the $p+q$ ordinary one-forms $\{\theta^{a}\}$ are linearly
independent on a $s=(p+q)-$dimensional sub-manifold $S\subseteq M$,
$(p+q)\leqq n$, so that they form an orthonormal basis for a metric of
signature $(p,q)$, $ds^{2}=\eta_{ab}\theta^{a}\otimes\theta^{b}$ on $S$.
\ There are four cases to consider on $S$.

Case (ia) where $\epsilon=\sigma=0$. \ 

In this case it follows from the first two of Eqs.(54) that $\omega_{b}^{a}$
is the Levi-Civita connection of the metric $ds^{2}$ and the metric is flat.
\ Hence coordinates $\{x^{a}\}$ can be introduced, and a gauge chosen so that
$\theta^{a}=dx^{a}$ and $\omega_{b}^{a}=0$ with the Levi-Civita covariant
derivative $\nabla_{a}$ now the partial derivative $\partial_{a}$. \ The last
two of Eqs.(54) then become, in terms of their components with respect to the
basis one-forms,
\begin{align}
\partial_{\lbrack d}\Theta_{bc]}^{a}  &  =\kappa_{\lbrack bcd]}^{a},\\
\partial_{\lbrack e}\kappa_{\mid b\mid cd]}^{a}  &  =0,\nonumber
\end{align}
where $\Theta^{a}=\frac{1}{2}\Theta_{bc}^{a}\theta^{b}\theta^{c}$,
$\Theta_{bc}^{a}=\Theta_{\lbrack bc]}^{a}$, and $\kappa_{b}^{a}=\frac{1}%
{2}\kappa_{bcd}^{a}\theta^{c}\theta^{d}$, $\kappa_{abcd}=$ $\kappa_{\lbrack
ab]cd}=$ $\kappa_{ab[cd]}$. \ When the additional condition
\begin{equation}
\Theta^{a}=0
\end{equation}
is imposed it follows that $\kappa_{bcd}^{a}$ has the symmetries of the
Riemann tensor and by an old local result due to Trautman, \cite{trautman} and
\cite{pirani}, the solution of these equations is then given by
\begin{equation}
\kappa_{bcd}^{a}=\eta^{ae}(\partial_{d}\partial_{b}\gamma_{ec}-\partial
_{b}\partial_{c}\gamma_{ed}-\partial_{e}\partial_{d}\gamma_{bc}+\partial
_{c}\partial_{e}\gamma_{bd}),
\end{equation}
where $\gamma_{ab}=\gamma_{ba}$ , and $\kappa_{bcd}^{a}$ are the components of
the (linearized) Riemann tensor of a metric $\eta_{ab}+2\gamma_{ab}$
linearized about flat space.

Case (ib) where $d\mathbf{m}=$ $\epsilon=0$ and $\sigma$ is non-zero.

In this case it follows from the first two of Eqs.(54) that $\omega_{b}^{a}$
is the Levi-Civita connection of the metric $ds^{2}$, and this metric has
constant curvature with Ricci scalar $R=\frac{\sigma}{l^{2}}s(s-1)$ when
$s>1$. \ The last two of Eqs. (54) become, in terms of their components with
respect to the basis one-forms,
\begin{align}
\nabla_{\lbrack d}\Theta_{bc]}^{a}  &  =\kappa_{\lbrack bcd]}^{a},\\
\nabla_{\lbrack e}\kappa_{\mid b\mid cd]}^{a}  &  =\frac{\sigma}{l^{2}}%
(\Theta_{b[de}\delta_{c]}^{a}-\eta_{b[c}\Theta_{de]}^{a}),\nonumber
\end{align}
where $\nabla$ denotes the Levi-Civita covariant derivative. \ In this case
when the\ additional conditions
\begin{equation}
\Theta^{a}=0,\kappa_{.bad}^{a}=0.
\end{equation}
are imposed it follows from Eqs.(58) that $\kappa_{bcd}^{a}$ has the algebraic
symmetries of a Weyl tensor and satisfies $\nabla_{\lbrack e}\kappa_{\mid
b\mid cd]}^{a}=0$. \ Since the metric $ds^{2}$ has constant curvature it is
conformally flat and so $ds^{2}=(\exp2\mu)ds_{F}^{2}$, for the appropriate
function $\mu$, where $ds_{F}^{2}$ is flat. \ In coordinates and a gauge such
that $ds_{F}^{2}=\eta_{ab}dx^{a}\otimes dx^{b}$, $\theta^{a}$ =$(\exp
\mu)dx^{a}$ and the one-forms $dx^{a}$\ constitute the orthonormal frame for
the flat metric, the flat metric Levi-Civita connection is zero and the
covariant exterior derivative for the flat metric is just the ordinary
exterior derivative. \ If $\kappa_{Fb}^{a}=(\exp\mu)\kappa_{b}^{a}$ then
$d\kappa_{Fb}^{a}=(\exp\mu)D\kappa_{b}^{a}$ and so $\partial_{\lbrack e}%
\kappa_{F\mid b\mid cd]}^{a}=$ $\kappa_{F[bcd]}^{a}=\kappa_{F.bad}^{a}$ $=0$.
Hence, using again Trautman's result mentioned above, $\kappa_{Fbcd}^{a}$ are
the components of the (linearized) Riemann tensor of the linearized metric
$\eta_{ab}+2\gamma_{ab}$. \ Consequently
\begin{equation}
\kappa_{bcd}^{a}=\exp(-3\mu)\eta^{ae}(\partial_{d}\partial_{b}\gamma
_{ec}-\partial_{b}\partial_{c}\gamma_{ed}-\partial_{e}\partial_{d}\gamma
_{bc}+\partial_{c}\partial_{e}\gamma_{bd})
\end{equation}
\ Further more, since $\kappa_{.bad}^{a}=\kappa_{F.bad}^{a}=0$, the linearized
metric components $\gamma_{ab}$ satisfy the linearized Einstein vacuum field
equations with zero cosmological constant.

In summary, if $\epsilon=0$ then the generalized connection, with $\Theta
^{a}=0,\kappa_{.bad}^{a}=0$, is flat if and only if the metric $ds^{2}$
$=\eta_{ab}\theta^{a}\otimes\theta^{b}$ has constant curvature so a coframe
can be chosen such that $\theta^{a}$ =$\exp(\mu)dx^{a}$ . $\ $Furthermore
$\kappa_{bcd}^{a}=\exp(-3\mu)\eta^{ae}(\partial_{d}\partial_{b}\gamma
_{ec}-\partial_{b}\partial_{c}\gamma_{ed}-\partial_{e}\partial_{d}\gamma
_{bc}+\partial_{c}\partial_{e}\gamma_{bd})$ where $\gamma_{ab}$ $=\gamma_{ba}$
satisfies the Einstein vacuum field equations, with zero cosmological
constant, linearized about flat space.

Case (iia) where $d\mathbf{m}=$ $\epsilon$ is non-zero but $\sigma$ is zero.

In this case the generalized connection is flat if and only if the connection
$\omega_{b}^{a}$ has torsion $\epsilon\Theta^{a}$ and\ curvature two-form
$\epsilon\kappa_{b}^{a}$. \ When Eqs.(59) are imposed so that the torsion
vanishes then the connection is the Levi-Civita connection of the metric
$ds^{2}$ and by the second of Eqs.(59) its Ricci tensor is zero.

Case (iib) where $d\mathbf{m}=$ $\epsilon$ and $\sigma$ are both non-zero.

In this case Eqs.(54) are all satisfied, and hence the generalized connection
is flat, if and only if the connection $\omega_{b}^{a}$ has torsion
$\epsilon\Theta^{a}$ and curvature with components%
\begin{align}
\Omega_{b}^{a}  &  =\frac{1}{2}R_{bcd}^{a}\theta^{c}\theta^{d},\\
R_{bcd}^{a}  &  =\epsilon\kappa_{bcd}^{a}+\frac{\sigma}{l^{2}}(\delta_{c}%
^{a}\eta_{bd}-\delta_{d}^{a}\eta_{bc}).\nonumber
\end{align}
When the additional conditions given in Eq.(59) are imposed the generalized
connection is flat if and only if $\omega_{b}^{a}$ is the Levi-Civita
connection of the metric which satisfies the full Einstein vacuum field
equations with cosmological constant given by $\frac{\sigma}{l^{2}}s(s-1)$.

The solutions of Eqs.(54) can be found by using Eq.(33) and the fact that the
flat connection is given by $\mathbf{A}=\mathbf{g}^{-1}d\mathbf{g}$ where
$\mathbf{g}\in G_{(1)}$. \ This is done in the following example in the cases
where $\sigma=0$. \ The other cases, i(b) and ii(b), where $\sigma$ is
non-zero can be treated similarly using the groups $G=SO(p+1,q)$ when
$\sigma=1$ and $G=SO(p,q+1)$ when $\sigma=-1$.

\textbf{Example 3: \ }Consider the cases i(a) and ii(a) above where $\sigma
=0$. \ In this case $G_{(1)}=ISO(p,q)_{(1)}$. \ The $(p+1,q+1)$ matrix
representation for $\mathbf{g=(}1\mathbf{+}h\mathbf{m)}g\in G_{(1)}$ can be
written in terms of the matrices%
\[
h\mathbf{=}\left(
\begin{array}
[c]{cc}%
h_{b}^{a} & 0\\
\frac{h_{b}}{l} & 0
\end{array}
\right)  ,\text{ }t=\left(
\begin{array}
[c]{cc}%
\delta_{b}^{a} & 0\\
\frac{\tau_{b}}{l} & 1
\end{array}
\right)  ,\text{ }g_{0}=\left(
\begin{array}
[c]{cc}%
L_{b}^{a} & 0\\
0 & 1
\end{array}
\right)  ,
\]
where, without loss of generality, $g=tg_{0}$. \ Here $h$ is an ordinary
one-form with values in the Lie algebra of $ISO(p,q)$, $g$ is an
$ISO(p,q)$-valued zero-form and the matrix-valued function $\left(  L_{b}%
^{a}\right)  $ takes values in $SO(p,q)$\ Working modulo $g_{0}$,
$\mathbf{A}=\mathbf{g}^{-1}d\mathbf{g}$ is
\[
\left(
\begin{array}
[c]{cc}%
-\epsilon h_{b}^{a}+(dh_{b}^{a}-\epsilon h_{c}^{a}h_{b}^{c})\mathbf{m} & 0\\
\frac{1}{l}\{d\tau_{b}+\epsilon\tau_{c}h_{b}^{c}-\epsilon h_{b}+[dh_{b}%
-\epsilon h_{c}h_{b}^{c}-\tau_{a}(dh_{b}^{a}-\epsilon h_{c}^{a}h_{b}%
^{c})]\mathbf{m\}} & 0
\end{array}
\right)  .
\]
and this expression together with Eqs.(44) and (48) can be used to evaluate
(modulo $g_{0}$ gauge transformations) $\omega_{b}^{a}$, $\kappa_{b}^{a},$
$\theta^{a}$ and $\Theta^{a}$,%
\begin{align*}
\theta^{a}  &  =d\tau^{a}-\epsilon\tau^{b}h_{b}^{a}-\epsilon h^{a},\text{
}\Theta^{a}=-\tau^{b}(dh_{b}^{a}-\epsilon h_{c}^{a}h_{b}^{c})-dh^{a}-\epsilon
h^{b}h_{b}^{a},\\
\omega_{b}^{a}  &  =-\epsilon h_{b}^{a},\text{ }\kappa_{b}^{a}=-(dh_{b}%
^{a}-\epsilon h_{c}^{a}h_{b}^{c}).
\end{align*}

The association of a linear problem with certain physically interesting
non-linear partial differential equations, including those admitting soliton
solutions, has enabled techniques such as the inverse scattering methods to be
used to solve them, see for example \cite{ablowitz} and \cite{mason}. \ These
linear problems often have a geometrical interpretation, as for example is the
case with the soliton connection and the equations of parallel transport of a
linear connection on a principal SL(2,R) bundle, \cite{crampin}. \ In this
regard it should be noted that the (linear) equations of parallel transport,
Eq.(31), for the generalized connections considered in this section have
integrabilty conditions which can include, as discussed above, the Einstein
vacuum field equations.

\section{Generalized Chern-Simons Lagrangians}

For the connection and curvature given by Eqs.(44), (46) and (47) the
generalized Chern-Pontrjagin and Chern-Simons forms are%
\begin{align}
\mathbf{CP}  &  \mathbf{=}\frac{1}{8\pi^{2}}[\mathbf{F}_{b}^{a}\mathbf{F}%
_{a}^{b}-2\sigma\mathbf{F}^{a}\mathbf{F}_{a}]\\
\mathbf{CS}  &  \mathbf{=}\frac{1}{8\pi^{2}}[\mathbf{A}_{b}^{a}\mathbf{F}%
_{a}^{b}-\frac{1}{3}\mathbf{A}_{b}^{a}\mathbf{A}_{c}^{b}\mathbf{A}_{a}%
^{c}-2\sigma\mathbf{A}^{a}\mathbf{F}_{a}+\sigma\mathbf{A}_{b}^{a}%
\mathbf{A}^{b}\mathbf{A}_{a}].\nonumber
\end{align}

Computation, using Eqs.(48) and (49), gives%
\begin{align}
\mathbf{CS}  &  \mathbf{=}CS_{0}+\frac{1}{8\pi^{2}}\{\frac{2\sigma}{l^{2}%
}(\epsilon\theta^{a}\Theta_{a}-\theta^{a}D\theta_{a})-\epsilon\omega_{.b}%
^{a}K_{.a}^{b}\\
&  +d(\omega_{.b}^{a}\kappa_{.a}^{b}-\frac{2\sigma}{l^{2}}\theta^{a}\Theta
_{a})\mathbf{m}\nonumber\\
&  +[\epsilon\kappa^{a}._{b}\kappa_{.a}^{b}-2\kappa_{.b}^{a}\Omega_{.a}%
^{b}+\frac{2\sigma}{l^{2}}(2\Theta^{a}D\theta_{a}-\kappa_{ab}\theta^{a}%
\theta^{b}-\epsilon\Theta^{a}\Theta_{a})]\mathbf{m\},}\nonumber
\end{align}
where $CS_{0}$ is the ordinary Chern-Simons $3-$form%
\begin{equation}
CS_{0}=\frac{1}{8\pi^{2}}\{\omega_{.b}^{a}\Omega_{a}^{b}.-\frac{1}{3}%
\omega_{.b}^{a}\omega_{.c}^{b}\omega_{.a}^{c}\}.
\end{equation}
When the polychain is itself a boundary with $\mathbf{c}_{3}=\partial
\mathbf{c}_{4}$ as in Eq.(41), then%
\begin{align}
\int_{\mathbf{c}_{3}}\mathbf{CS}  &  \mathbf{=}\frac{1}{8\pi^{2}}%
\{\int_{\partial c_{4}}[8\pi^{2}CS_{0}-\frac{2\sigma}{l^{2}}\theta^{a}%
D\theta_{a}]\\
&  +\epsilon\int_{c_{4}}[\epsilon\kappa_{.b}^{a}\kappa_{.a}^{b}-2\kappa
_{.b}^{a}\Omega_{.a}^{b}+\frac{2\sigma}{l^{2}}(2\Theta^{a}D\theta_{a}%
-\kappa_{ab}\theta^{a}\theta^{b}-\epsilon\Theta^{a}\Theta_{a})]\nonumber\\
&  +\int_{\partial c_{5}}[\epsilon\kappa_{.b}^{a}\kappa_{.a}^{b}-2\kappa
_{.b}^{a}\Omega_{.a}^{b}+\frac{2\sigma}{l^{2}}(2\Theta^{a}D\theta_{a}%
-\kappa_{ab}\theta^{a}\theta^{b}-\epsilon\Theta^{a}\Theta_{a})]\}.\nonumber
\end{align}
Next consider this expression as a generalized Chern-Simons action integral.
\ Computing the variation of the 3-form $\mathbf{CS}$ and this generalized
Chern-Simons action gives
\begin{align}
\delta\mathbf{CS}  &  \mathbf{=}\frac{1}{8\pi^{2}}\{\frac{2\sigma}{l^{2}%
}\delta\theta^{a}(\epsilon\Theta_{a}-2D\theta_{a})+\frac{2\sigma\epsilon
}{l^{2}}\mathbf{\delta\Theta^{a}\theta_{a}+\delta\omega}_{b}^{a}(2\Omega
_{.a}^{b}-\epsilon\kappa_{a}^{b}+\frac{2\sigma}{l^{2}}\theta_{a}\theta
^{b})-\epsilon\delta\kappa_{.b}^{a}\omega_{.a}^{b}\\
&  +d(\delta\omega_{.b}^{a}\omega_{.a}^{b}-\frac{2\sigma}{l^{2}}\delta
\theta^{a}\theta_{a})+d(\frac{2\sigma}{l^{2}}\delta\theta^{a}\Theta_{a}%
-\frac{2\sigma}{l^{2}}\delta\Theta^{a}\theta_{a}-\delta\omega_{.b}^{a}%
\kappa_{.a}^{b}+\delta\kappa_{.b}^{a}\omega_{.a}^{b})\mathbf{m}\nonumber\\
&  +[\frac{4\sigma}{l^{2}}\delta\theta^{a}(D\Theta_{a}-\theta^{b}\kappa
_{ab})+\frac{4\sigma}{l^{2}}\delta\Theta^{a}(D\theta_{a}-\epsilon\Theta
_{a})+2\delta\omega_{.b}^{a}(\frac{2\sigma}{l^{2}}\theta^{b}\Theta_{a}%
-D\kappa_{.a}^{b})\nonumber\\
&  +2\delta\kappa_{b}^{a}(\epsilon\kappa_{.a}^{b}-\Omega_{.a}^{b}-\frac
{\sigma}{l^{2}}\theta_{a}\theta^{b})]\mathbf{m\}}.\nonumber
\end{align}%
\begin{align}
\delta\int_{\mathbf{c}_{3}}\mathbf{CS}  &  \mathbf{=}\frac{1}{8\pi^{2}}%
\{\int_{\partial c_{4}}[\frac{4\sigma}{l^{2}}\delta\theta^{a}(\epsilon
\Theta_{a}-D\theta_{a})\mathbf{+2\delta\omega}_{.b}^{a}(\Omega_{.a}%
^{b}-\epsilon\kappa_{.a}^{b}+\frac{\sigma}{l^{2}}\theta_{a}\theta^{b})]\\
&  +\int_{\partial c_{5}+\epsilon c_{4}}[\frac{4\sigma}{l^{2}}\delta\theta
^{a}(D\Theta_{a}-\theta^{b}\kappa_{ab})+\frac{4\sigma}{l^{2}}\delta\Theta
^{a}(D\theta_{a}-\epsilon\Theta_{a})\nonumber\\
&  +2\delta\omega_{.b}^{a}(\frac{\sigma}{l^{2}}\theta^{b}\Theta_{a}%
-\frac{\sigma}{l^{2}}\theta_{a}\Theta^{b}-D\kappa_{.a}^{b})+2\delta\kappa
_{b}^{a}(\epsilon\kappa_{.a}^{b}-\Omega_{.a}^{b}-\frac{\sigma}{l^{2}}%
\theta_{a}\theta^{b})]\}.\nonumber
\end{align}

When $\sigma=1$ so that the gauge group is $SO(p+1,q),$ or $\sigma=-1$ so that
the gauge group is $SO(p,q+1)$, the variation of the Chern-Simons integral
vanishes for arbitrary variations of the forms $\theta^{a},\Theta^{a}%
,\omega_{.b}^{a}$ and $\kappa_{b}^{a}$ if and only if the variational
equations on $(\partial c_{5}+\epsilon c_{4})$ are given by the previously
discussed Eq.(54), that is they correspond to the vanishing of the generalized
curvature $\mathbf{F}$ there.

It should be noted that if the generalized connection $\mathbf{A}$ in the
generalized gauge in which $\Theta^{a}=0$ had been used only the last three of
the above variational equations (with $\Theta^{a}=0$) would have been obtained
from the variational principle. \ However these imply, via the Bianchi
identity, the first, that is $D\theta^{a}=0$.

Finally consider the case where $\sigma$ is zero and the gauge group is
$ISO(p,q)$. \ Now only the equations
\begin{align}
\Omega_{b}^{a}  &  =\epsilon\kappa_{b}^{a},\\
D\kappa_{.b}^{a}  &  =0\nonumber
\end{align}
follow from the action principle on $(\partial c_{5}+\epsilon c_{4})$. \ Then
in case (i), where $\epsilon=0,$ these variational equations reduce to
$\Omega_{b}^{a}=0$ and $D\kappa_{.b}^{a}=0$. \ In case (ii), where $\epsilon$
is non-zero, Eq.(68) merely identifies $\Omega_{b}^{a}$ with $\epsilon
\kappa_{b}^{a}$ on $(\partial c_{5}+\epsilon c_{4})$.

Similarly the equations arising from non vanishing variations on $\partial
c_{4}$ can be read off Eq.(67).

The variational equations and generalized gauge freedom change when
constraints are imposed on either $\Theta^{a}$ or $\kappa_{b}^{a}$ and this
will be illustrated in the next section.

\section{Actions and Lorentzian four-metrics}

In this section an action and gravitational field equations for Lorentzian
metrics and four dimensional space-times are considered. \ In the equations of
the previous section the choice $p=3,q=1$ is made so that the four-metric
$\eta_{ab}$ has Lorentzian signature. \ Furthermore the two-forms $\kappa
_{b}^{a}$ in Eq. (48) and subsequent equations are taken to be of the form
\begin{equation}
\kappa_{b}^{a}=\kappa\varepsilon_{bcd}^{a}\theta^{c}\theta^{d}%
\end{equation}
where $\varepsilon_{abcd}$ is the Levi-Civita symbol with $\varepsilon
_{1234}=1$, and $\kappa$ is a non-zero constant, see also \cite{rob6}. \ Hence
the generalized connections considered here are given by Eqs.(44),(48) and
(69) with $p+q+1=5$. \ This restricts the variations considered in the
previous section and the (gauge) structure group is reduced to $SO(3,1)$.

With this choice of $\kappa_{b}^{a}$\ the Chern-Simons action integral,
Eq.(65), becomes%
\begin{align}
\int_{\mathbf{c}_{3}}\mathbf{CS}  &  \mathbf{=}\frac{1}{8\pi^{2}}%
\{\int_{\partial c_{4}}[8\pi^{2}CS_{0}-\frac{2\sigma}{l^{2}}\theta^{a}%
D\theta_{a}]\\
&  +\int_{\epsilon c_{4}+\partial c_{5}}[-2\kappa\epsilon_{.bcd}^{a}\theta
^{c}\theta^{d}\Omega_{.a}^{b}-\frac{2\sigma\kappa}{l^{2}}\epsilon_{abcd}%
\theta^{a}\theta^{b}\theta^{c}\theta^{d}+\frac{2\sigma}{l^{2}}(2\Theta
^{a}D\theta_{a}-\epsilon\Theta^{a}\Theta_{a})].\nonumber
\end{align}
with variation%
\begin{align}
\delta\int_{\mathbf{c}_{3}}\mathbf{CS}  &  \mathbf{=}\frac{1}{8\pi^{2}}%
\{\int_{\partial c_{4}}[\frac{4\sigma}{l^{2}}\delta\theta^{a}(\epsilon
\Theta_{a}-D\theta_{a})\mathbf{+2\delta\omega}_{.b}^{a}(\Omega_{.a}%
^{b}-\epsilon\kappa\epsilon_{.acd}^{b}\theta^{c}\theta^{d}+\frac{\sigma}%
{l^{2}}\theta_{a}\theta^{b})]\\
&  +\int_{\partial c_{5}+\epsilon c_{4}}[4\delta\theta^{a}(\frac{\sigma}%
{l^{2}}D\Theta_{a}-\frac{2\kappa\sigma}{l^{2}}\kappa\epsilon_{abcd}\theta
^{c}\theta^{d}-\kappa\epsilon_{a.cd}^{b}\Omega_{.b}^{c}\theta^{d})\nonumber\\
&  +\frac{4\sigma}{l^{2}}\delta\Theta^{a}(D\theta_{a}-\epsilon\Theta
_{a})+2\delta\omega_{.b}^{a}(\frac{\sigma}{l^{2}}\theta^{b}\Theta_{a}%
-\frac{\sigma}{l^{2}}\theta_{a}\Theta^{b}-\kappa\epsilon_{.acd}^{b}%
D(\theta^{c}\theta^{d}))]\}.\nonumber
\end{align}

The Euler-Lagrange equations now follow from Eq.(71) and the variation of the
independent variables $\theta^{a}$, $\Theta^{a}$ and\ $\omega_{b}^{a}$ . \ On
$\epsilon c_{4}+\partial c_{5\text{ }}$, taken to be a four dimensional
manifold $S$, they are
\begin{align}
\frac{\sigma}{l^{2}}(D\Theta^{a}-2\kappa\epsilon_{.bcd}^{a}\theta^{b}%
\theta^{c}\theta^{d})+\kappa\epsilon_{.bcd}^{a}\Omega^{bc}\theta^{d}  &  =0,\\
\frac{\sigma}{l^{2}}(D\theta^{a}-\epsilon\Theta^{a})  &  =0,\nonumber\\
2\kappa D(\theta^{a}\theta^{b})+\frac{\sigma}{l^{2}}\epsilon_{..cd}^{ab}%
\theta^{c}\Theta^{d}  &  =0.\nonumber
\end{align}
When it is assumed that \ the one-forms $\{\theta^{a}\}$ are linearly
independent on $S$ and hence define a Lorentzian four-metric $ds^{2}=\eta
_{ab}\theta^{a}\otimes\theta^{b}$ on $S$ it follows from these equations that,
whether the constant $\epsilon$ is zero or not, $\Theta^{a}=0$, $\omega
_{b}^{a}$ is the Levi-Civita connection of the metric with curvature
$\Omega_{b}^{a}$ and Einstein's vacuum field equations $G_{ab}=-6\frac{\sigma
}{l^{2}}\eta_{ab}$, where $G_{ab}$ are the components of the Einstein tensor
$R_{acb}^{c}-\frac{1}{2}\eta_{ab}R_{..cd}^{cd}$, are satisfied in each of the
three cases $\sigma=\pm1$ or $\sigma=0.$

Further equations will arise from non-vanishing variations on $\partial
c_{4}=\partial S$ and can be read off Eq.(71). \ When $\kappa_{b}^{a}$ is
given by Eq.(69) the vanishing of the generalized curvature implies that the
four one-forms $\{\theta^{a}\}$ are not linearly independent except when
$\epsilon=\sigma=0.$ \ In this case flatness implies that, when the one-forms
are linearly independent and define a Lorentz metric on a four-manifold, this
metric is flat, with $\omega_{b}^{a}$ being the flat Levi-Civita connection of
the metric and with the two-forms$\{\Theta^{a}\}$ satisfying the equations
$D\Theta^{a}=\kappa\epsilon_{bcd}^{a}\theta^{b}\theta^{c}\theta^{d}$.

In conclusion it should be noted that interesting observations on symmetry
breaking and gravitational actions, from the point of view of ordinary Cartan
connections and Cartan geometry, can be found in \cite{wise} and an approach
to general relativity which is different from, but has aspects in common with,
this work can be found in \cite{mikovic}.

\end{document}